\documentclass[12pt]{iopart}

\usepackage[dvips]{graphicx}

\eqnobysec
\newcommand{\beq}{\begin{equation}}
\newcommand{\beqa}{\begin{eqnarray}}
\newcommand{\eeq}{\end{equation}}
\newcommand{\eeqa}{\end{eqnarray}}

\renewcommand{\c}{{\cal C}}

\newcommand{\s}{{\sigma}}

\newcommand{\w}{{\bar w}}

\begin{document}

\title{Nonequilibrium Stationary States and Phase Transitions in Directed Ising Models}
\date{\today}
\author{Claude Godr\`eche$^1$, Alan \ J.\ Bray$^2$}
\address{
Institut de Physique Th\'eorique, CEA Saclay,
91191 Gif-sur-Yvette cedex, France}
\address{$^2$School of Physics and Astronomy, University of Manchester,
Manchester M13 9PL, U.K.}
\begin{abstract}

We study the nonequilibrium properties of directed Ising models with non conserved dynamics, 
in which each spin is influenced by only a subset of its nearest neighbours. 
We treat the following models: (i) the one-dimensional chain; (ii) the two-dimensional square lattice; (iii) the two-dimensional triangular lattice;
(iv) the three-dimensional cubic lattice.
We raise and answer the question: (a) Under what conditions is the stationary state described by the equilibrium Boltzmann-Gibbs distribution?
We show that for models (i), (ii), and (iii), in which each spin ``sees'' only half of its neighbours, there is a unique set of transition rates, namely with exponential dependence in the local field, for which this is the case. 
For model (iv), we find that
any rates satisfying the constraints required for the stationary measure to be Gibbsian should satisfy detailed balance, ruling out the possibility of directed dynamics.
We finally show that directed models on lattices of coordination number $z\ge8$ with exponential rates cannot accommodate a Gibbsian stationary state.
We conjecture that this property extends to any form of the rates. We are thus led to the conclusion that directed models with Gibbsian stationary states only exist in dimension one and two.
We then raise the question: (b) Do directed Ising models, augmented 
by Glauber dynamics, exhibit a phase transition to a ferromagnetic state? 
For the models considered above, the answers are open problems, to the exception of the 
simple cases (i) and (ii). For Cayley trees, 
where each spin sees only the spins further from the root, we show that there is a phase transition provided the branching ratio, $q$, satisfies $q \ge 3$. 

\end{abstract}

\maketitle

\section{Introduction}
A long-standing theme of the statistical physics of nonequilibrium systems is the question of the 
nature of their stationary state, and in particular of the existence
of phase transitions at stationarity.
By nonequilibrium systems, we mean systems whose dynamics is non reversible, i.e., such that the
dynamical rules defining the model do not obey detailed balance.
For example, systems which are driven, e.g. submitted to a field, do not obey detailed balance.
The nature of the stationary state of driven systems has been well investigated for a number of models, among which
are the Zero Range Process~\cite{zrp}, the Asymmetric Simple Exclusion Process~\cite{asep}, or the KLS model~\cite{kls,gl}.

In contrast, the same questions have not yet been addressed in a comprehensive way for directed Ising models with non conserved dynamics.
In one dimension, for example, a spin is
influenced only by its right neighbour, while for a 
two-dimensional
square lattice, for example, each spin is influenced only by the neighbouring spins
above (``North'') and to the right (``East''), and not by the spins
below and to the left. The asymmetry of the interactions means that
the detailed balance condition is not obeyed, and therefore there is no conventional 
equilibrium. Again one speaks instead in terms of a stationary state.
However, the latter differs from the stationary state that prevails for driven systems in that it does not carry a macroscopic current.
In the present work we endeavour a study of these questions, namely the nature of the stationary state measure and the existence of phase transitions, for directed Ising models with single spin-flip dynamics.

Usually, when the dynamics of a model does not fulfill detailed
balance, one expects its stationary measure to be
non-Gibbsian. Consider for instance the Ising chain with conserved
dynamics, and totally asymmetric rules: only the $+-$ bond 
is updated, not the $-+$ bond.
As pointed by Katz, Lebowitz and Spohn~\cite{kls}, while generically 
the stationary measure of this model is
non-Gibbsian, for special
choices of the rates defining the dynamics, the stationary measure is the same as that of the model with symmetric dynamics, i.e., the equilibrium Boltzmann-Gibbs distribution. 
The one-dimensional KLS model, which can also be viewed as the 
hopping of one species of particles with exclusion on a lattice, has 
been further analysed and generalised to the case of two species 
of particles hopping with exclusion~\cite{gl}, where, again, the stationary measure is the same
whether the system is driven or not.
To date, for spin models with non-conserved dynamics, no systematic analysis of
this kind has been performed. The main purpose of this paper is to provide such an analysis, for the Ising model, in one, two, and three dimensions. 

We treat the following models: (i) the one-dimensional chain; (ii) the two-dimensional square lattice; (iii) the two-dimensional triangular lattice;
(iv) the three-dimensional cubic lattice.
The question to
be answered is how to choose the transition rates, for the single spin-flip dynamics, in such a way
that they lead to a stationary state with the equilibrium measure
(whether or not the dynamics is directed)? The method
consists in writing the master equation at stationarity, then imposing 
the equilibrium measure~\cite{gl}. One thus finds a set of constraint equations
on the rates. Additional
constraints come from the choice of dynamics (with
or without spin symmetry, directed or not).
For the undirected models (i.e., with symmetric dynamics) we find, as expected, 
that any rates leading to the Gibbs measure satisfy detailed balance. 
For the directed case, we show that for models (i), (ii), and (iii), in which each spin ``sees'' only half of its neighbours, there is a unique set of transition rates, namely with exponential dependence in the local field, for which this is the case. 
By local field we mean the restricted field felt by the flipping spin.
For a variant of model (ii), where each spin sees only its West, North and East neighbours, the property still holds but the set of rates have no longer a simple dependence in the local field.
For model (iv), we find that
any rates satisfying the constraints required for the stationary measure to be Gibbsian should satisfy detailed balance, ruling out the possibility of directed dynamics.
We finally show that directed models on lattices of coordination number $z\ge8$ with exponential rates cannot accommodate a Gibbsian stationary state.
We conjecture that this property extends to any form of the rates. We are thus led to the conclusion that directed models with Gibbsian stationary states only exist in dimension one and two.

The second purpose of this paper is to investigate
whether directed Ising models, augmented 
by Glauber dynamics, can exhibit a phase transition to a ferromagnetic state.
This question was motivated by a recent paper~\cite{stau}, where Lima and Stauffer considered Ising models
with directed interactions in two to five space dimensions, evolving via 
Glauber dynamics
\cite{glau}. 
They find in particular that for the two-dimensional square lattice
where each spin is influenced only by the North and East neighbouring spins
there is no ferromagnetic phase transition, in contrast to the usual equilibrium Ising model which
has a second-order phase transition in two dimensions. 
In the present work we show how the absence of a phase transition arises
both for the directed Ising chain and for the two-dimensional directed square lattice
considered in~\cite{stau}.
We then investigate Cayley trees, with the
influence directed from the tips towards the root, and show that there
is a phase transition for coordination number $z \ge 4$ but not for
$z=3$, i.e., for branching ratio $q \ge 3$, since $q=z-1$. 

The outline of the paper is as follows. We begin, in the following
section, by giving an extensive analysis of the one-dimensional case.
We show in particular that, for the directed Ising chain, there is a unique set of rates which produce stationary states described by the Gibbs measure. 
We then address in section~\ref{regular} the case of two-dimensional and three-dimensional Ising models on regular lattices.
Section~\ref{coordi} is devoted to another approach,
which allows to determine which lattices, with given coordination
number, can sustain directed models with rates of exponential form.
We then proceed, in section~\ref{glob},
by studying directed models with Glauber dynamics. 
We show
that for the two-dimensional square lattice there is no phase transition to a
ferromagnetic state, the latter result being in agreement with the
numerical data of~\cite{stau}. The same result holds for the directed Ising chain.
We show that for
the Cayley tree model, a phase transition occurs if the branching
ratio is greater than or equal to three. 
Some generalizations are given in the 
Appendix. 
Section~\ref{concl} concludes with a discussion and summary of the results.


\section{Equilibrium Measures for Nonequilibrium Stationary States}
\label{chain}
In this section we pose the question: ``Under what conditions are 
the stationary states of Ising models described by the 
Boltzmann-Gibbs distribution?''. 
We start by the one-dimensional Ising model. 
We consider higher dimensional models in sections~\ref{regular} and~\ref{coordi}.

\subsection{The method illustrated on the case of the Ising chain}


We derive a set of constraints between the transition rates that need
to be satisfied in order for the stationary state measure to be Gibbsian. 
For symmetric dynamics, the constraints are
satisfied if and only if the rates satisfy detailed balance. For
directed dynamics, however, the constraints uniquely determine the
rates (up to an overall timescale). 

The energy of configuration $\c=\{\s_1,\ldots,\s_{i},\ldots,\s_{N}\}$ is given by
\beq\label{erg1d}
E(\c)=-J\sum_{i}\sigma_i\sigma_{i+1},
\eeq
and we choose periodic boundary conditions.
The dynamics consist in flipping a spin, chosen at random, say spin $i$, with a rate 
$w(\c_i|\c)$ corresponding to the transition between configurations $\c$ and 
$\c_i=\{\s_1,\ldots,-\s_{i},\ldots,\s_{N}\}$.
The change in energy due to the flip reads
\beq\label{del}
\Delta E=E(\c_i)-E(\c)=2\sigma_i h_i=2\s_i\,J(\s_{i-1}+\s_{i+1}),
\eeq
where we denote the local field acting on spin $i$ by $h_i$.
At stationarity, the master equation expresses that losses are equal to gains, and reads
\beq\label{master0}
P(\c)\sum_{i}w(\c_i|\c)=\sum_{i}w(\c|\c_i)P(\c_i),
\eeq
where, by hypothesis,
\beq
P(\c)\propto\e^{-\beta E(\c)}.
\eeq
After division of both sides by the weight $P(\cal C)$,
eq.~(\ref{master0}) can be rewritten as
\beq\label{mast}
\sum_{i}w(\c_i|\c)-w(\c|\c_i)\e^{-\beta\Delta E}=0,
\eeq
which is also, in the present context, the general form of the stationary master equation in any dimension.

\begin{table}[ht]
\caption{List of moves and corresponding energy difference $\Delta E$ (eq.~(\ref{del})) for the Ising chain.}
\label{2d}
\begin{center}
\begin{tabular}{|c|c|c||c|c|c|}
\hline
Move&Rate&$\Delta E$&Move&Rate&$\Delta E$\\
\hline
$+++\to +-+$&$w_{++}$&$4J$&
$+-+\to +++$&$\w_{++}$&$-4J$\\
$++-\to +--$&$w_{+-}$&$0$&
$+--\to ++-$&$\w_{+-}$&$0$\\
$-++\to --+$&$w_{-+}$&$0$&
$--+\to -++$&$\w_{-+}$&$0$\\
$-+-\to ---$&$w_{--}$&$-4J$&
$---\to -+-$&$\w_{--}$&$4J$\\
\hline
\end{tabular}
\end{center}
\end{table}
There are eight ($2^3$) possible rates $w(\c_i|\c)$ denoted by $w(\sigma_{i-1}\,\s_i\,\sigma_{i+1})$, or for short by 
$w_{\sigma_{i-1}\,\sigma_{i+1}}$ if $\sigma_i=+1$, 
or by $\w_{\sigma_{i-1}\,\sigma_{i+1}}$ if $\sigma_i=-1$ (see Table), corresponding to the 
flipping of spin $\sigma_i$ for the eight possible {\it motifs} (in the present case, the triplets appearing in the Table):
\begin{equation}
\sigma_{i-1} \,\sigma_{i} \,\sigma_{i+1}\rightarrow\sigma_{i-1}\,(- \sigma_{i}) \,\sigma_{i+1}.
\end{equation}

Let $N_{\sigma_{i-1} \,\sigma_{i} \,\sigma_{i+1}}$ be the number of such motifs in configuration $\c$, now considered as fixed. The balance equation~(\ref{mast}) is finally recast as
\begin{eqnarray}\label{master}
N_{+++}(w_{++}-\e^{-4 K}\,\w_{++})+N_{++-}(w_{+-}-\w_{+-})+N_{-++}(w_{-+}-\w_{-+})\nonumber\\
+N_{-+-}(w_{--}-\e^{4 K}\,\w_{--})+N_{+-+}(\w_{++}-\e^{4 K}\, w_{++})+N_{+--}(\w_{+-}- w_{+-})\nonumber\\
+N_{--+}(\w_{-+}-w_{-+})+N_{---}(\w_{--}-\e^{-4 K}\,w_{--})
=0,
\end{eqnarray}
where we denoted the reduced coupling constant $\beta J$ by $ K$.
A shorter notation for the same equation is
\beq\label{short}
\sum_{\alpha=1}^8 N(\mu_\alpha)\left[w(\mu_\alpha)-w(\bar\mu_\alpha)\e^{-\beta (E(\bar\mu_\alpha)-E(\mu_\alpha))}\right],
\eeq
where $N(\mu_\alpha)$ is the number of occurrences of the motif $\mu_\alpha$ in configuration $\c$, and 
$\bar \mu_\alpha$ is the motif obtained after flipping the central spin in the motif.

Equation~(\ref{short}) should be satisfied for any given configuration $\c$.
Since the quantities inside the brackets are couples of rates related by detailed balance conditions,
the left side of~(\ref{short}) vanishes identically as soon as these conditions are imposed on the rates.
However imposing detailed balance is not {\it a priori necessarily} the only way of solving~(\ref{short}), because the $N(\mu_\alpha)$ are not independent quantities.
One should therefore express the latter on a basis of independent quantities, in the present case correlators defined as follows. Let us denote by $I_i=(1+\s_i)/2$ the indicator variable for the presence of a $+$ spin on site $i$. Then, for example,
\beqa
N_{++-}=\sum_{i=1}^N I_{i}\,I_{i+1}(1-I_{i+2})=
\frac{1}{8}\sum_{i=1}^N (1+\sigma_{i})(1+\sigma_{i+1})(1-\sigma_{i+2})
\nonumber\\
=\frac{N}{8}(1+c_{1}+c_{2}-c_{3}+c_{12}-c_{13}-c_{23}-c_{123}),
\eeqa
where, for example,
\beq
c_1=\frac{1}{N}\sum_{i=1}^N \s_i, \quad c_2=\frac{1}{N}\sum_{i=1}^N \s_{i+1}, \quad c_{12}=\frac{1}{N}\sum_{i=1}^N \s_i\s_{i+1},
\eeq
and so on.
Translational invariance imposes the following identities between correlators,
\beq
c_1=c_2=c_3,\qquad c_{23}=c_{12}.
\eeq
Rewriting the master equation~(\ref{short}) in terms of the remaining independent correlators yields 
\beq
e_1+c_1\,e_2+c_{12}\,e_3+c_{13}\,e_4+c_{123}\,e_5=0,
\eeq
where the coefficients $e_1, \ldots, e_5$, are linear combinations of the expressions appearing in the brackets of eq.~(\ref{short}). This equation should be satisfied for any configuration $\c$, hence the coefficients of the correlators must vanish identically.
This provides five constraint equations, $e_1=\ldots=e_5=0$, not all independent, on the eight rates $w_{\s\s}$ and $\w_{\s\s}$.
Reducing this set of equations yields the final constraint equations:
\beqa
\w_{++}=\e^{4 K}\, w_{++}\label{db1}\\
\w_{--}=\e^{-4 K}\,w_{--}\label{db2}\\
\w_{+-}+\w_{-+}=w_{+-}+w_{-+}.\label{diff}
\eeqa
The first two equations are detailed balance conditions. They involve rates for moves corresponding to a non-zero value of the local field $ J(\sigma_{i-1}+\sigma_{i+1})=\pm2 J$. The third equation involves the rates for moves that do not imply a change in the energy, i.e. with zero local field (motion of a domain wall). 

The eight rates fulfilling the constraints~(\ref{db1}), (\ref{db2}, (\ref{diff}) therefore depend on five independent parameters. 
A general expression of these rates is
\beqa\label{gener1}
w(\sigma_{1}\,\s_2\,\sigma_{3})=
\frac{\alpha}{2}\Bigg\{1+\lambda \sigma_{1}-\lambda' \sigma_{2}+\left(2\frac{\lambda'}{\gamma}-\lambda\right)\sigma_3
+\delta\,\s_{1}\s_{3}+\epsilon\,\s_{1}\s_{2}\nonumber\\
-(\gamma(1+\delta)+\epsilon)\,\s_{2}\s_{3}-\lambda'\sigma_1\sigma_2\sigma_3
\Bigg\},
\eeqa
where $\lambda$, $\lambda'$, $\alpha$, $\delta$, $\epsilon$ are the five independent parameters,
while $\gamma$ is given by:
\beq
\gamma=\frac{\e^{4 K}-1}{\e^{4 K}+1}=\tanh 2K,
\eeq
and $\sigma_1,\sigma_2,\sigma_3$ are respectively the left, central, and right spins.
These rates correspond to the generic situation where the central spin is influenced unequally by the left and right spins.

Hereafter we restrict the discussion to the case of rates invariant by spin symmetry,
that is
\beq\label{spsym}
w_{\s\s}=\w_{-\s-\s}.
\eeq
In this case, the constraints~(\ref{db1}), (\ref{db2}, (\ref{diff}) reduce to a single equation,
\beq\label{seul}
w_{--}=\e^{4 K}\, w_{++},
\eeq
hence three rates remain unknown, namely $w_{++}, w_{+-}, w_{-+}$.
Setting $\lambda=\lambda'=0$, in order to satisfy~(\ref{spsym})
we obtain
\beqa\label{gener}
w(\sigma_{1}\,\s_2\,\sigma_{3})=
\frac{\alpha}{2}\left(1+\delta\,\s_{1}\s_{3}+\epsilon\,\s_{1}\s_{2}
-(\gamma(1+\delta)+\epsilon)\,\s_{2}\s_{3}
\right),
\eeqa
depending now on the three independent parameters, $\alpha$, $\delta$, $\epsilon$.

We now address two extreme cases of interest, where the dynamics is symmetric, then when it is totally asymmetric, or directed.

\subsection{Symmetric dynamics}

The dynamics is spatially symmetric, or undirected, if the left and right spins have an equal influence on the central spin. 
This right-left symmetry imposes therefore that
\beq\label{sym}
w_{+-}=w_{-+}.
\eeq
This fixes one of the remaining unknown rates.
By (\ref{sym}) we obtain $2 \epsilon+\gamma(1+\delta)=0$, and therefore
\beq
w(\sigma_{1}\,\s_2\,\sigma_{3})=
\frac{\alpha}{2}\left(1+\delta\,\s_{1}\s_{3}
-\frac{1}{2}\gamma(1+\delta)\,\s_{2}(\s_{1}+\s_{3})
\right),
\eeq
which is the most general form of the rates for symmetric dynamics, originally proposed by Glauber (see Appendix of~\cite{glau}), which now depend on two parameters.
This form satisfies the detailed balance conditions~(\ref{seul}) and (\ref{sym}).

Simplified forms of the rates are obtained by fixing the parameters $\alpha$ and $\delta$.
Examples of possible choices are the Glauber (or heat bath) rule:
\beq\label{heat}
w({\sigma_{ 1} \,\sigma_{2} \,\sigma_{3}})=
\frac{1}{2}\left(1-\s_2\tanh( K(\s_{ 1}+\s_{3}))\right)
=\frac{1}{\e^{\beta\Delta E}+1},
\eeq
or the Metropolis rule:
\beq
w({\sigma_{1} \,\sigma_{2} \,\sigma_{3}})=\min\left(1,\e^{-\beta\Delta E}\right),
\eeq
or yet the following rule:
\beq\label{exp}
w({\sigma_{1} \,\sigma_{2} \,\sigma_{3}})=
\e^{-\beta\,\Delta E/2}=\e^{-K\s_2 (\s_{ 1}+\s_{3})}.
\eeq
where $\Delta E$, eq.~(\ref{del}), is the change in energy of the system due to the flip.

\subsection{Directed dynamics} 

Let us now consider the case where the dynamics is totally directed, i.e., totally asymmetric. Assume for instance that 
spin $\sigma_i$ only looks to the right, hence
that the rates only depend on the right neighbour of the flipping spin:
\beq
 w_{++}=w_{-+},\quad w_{+-}=w_{--}.
\eeq
Carrying these conditions into (\ref{gener}), we obtain $\delta=\epsilon=0$,
and therefore
\beq\label{rate:direct}
w(\sigma_{ 1}\,\s_2\,\sigma_{3})=
\frac{\alpha}{2}\left(1
-\gamma\,\s_{2}\s_{3}
\right).
\eeq
This expression can be equivalently obtained by suppressing all terms where the spin $\sigma_1$ appears in the general expression of the rates~(\ref{gener1}).
We thus find a {\it unique} solution to the problem posed, up to the global time scale $\alpha$.
Fixing this scale by the choice $\alpha=2\cosh 2 K$, we obtain the compact form
\beq\label{ku1D}
w({\sigma_{1} \,\sigma_{2} \,\sigma_{3}})=\e^{-2 K\s_2\s_{3}},
\eeq
while, with the choice $\alpha=1$, we obtain the alternate form
\beq\label{alternate}
w({\sigma_{1} \,\sigma_{2} \,\sigma_{3}})=
\frac{1}{2}\left(1-\s_2\tanh( 2K \s_{3})\right).
\eeq

The directed dynamics is more constrained than the undirected one, as previously examplified for conserved dynamics of two or three species of particles in~\cite{gl}. No kinetically constraint models (i.e., with vanishing rates) can be devised in this case, in contrast with the case of the undirected dynamics which allows more freedom.

Let us finally mention that, as a special case of the generic expression~(\ref{gener}), where the central spin is unequally influenced by its neighbours, the following form of the rates has the virtue of interpolating between the undirected case~(\ref{exp}), and directed case~(\ref{ku1D}):
\beq
w({\sigma_{1} \,\sigma_{2} \,\sigma_{3}})=
\e^{-2 K\s_2 (x\s_{ 1}+(1-x)\s_{3})}, 
\eeq
with $0\le x\le1$. %

\section{Ising models on regular lattices}
\label{regular}

In this section we consider two- and three-dimensional Ising models on regular lattices.
The energy of configuration $\c=\{\s_1,\ldots,\s_{i},\ldots,\s_{N}\}$ is now given by
\beqa
E(\c)=- J\sum_{(i,j)}\s_i\s_j,
\eeqa
where $(i,j)$ are nearest neighbours. 
We follow step by step the method used in the previous section
for the determination of the transition rates leading to a stationary measure given by the Boltzmann-Gibbs distribution.
We start by the two-dimensional Ising model on the square lattice.
We proceed with the case of the two-dimensional triangular lattice, then of
the three-dimensional cubic lattice.

\subsection{Constraint equations on the rates for the square lattice}
On the square lattice each spin has four neighbours.
The change in energy when flipping the central spin, denoted by $\s$, reads
\beq
\Delta E=2\sigma J(\sigma_E+\sigma_N+\sigma_W+\sigma_S),
\eeq
where $\s_E$ is the East spin, $\s_N$ the North spin, etc.\footnote{The notation $\s_N$, where $N$ stands for North, should not be confused with the notation for the spin with index $N$, size of the system.}
The stationary master equation can be compactly written as in~(\ref{short}) with rates corresponding to the 32 motifs
$\s\s_E\s_N\s_W\s_S$. We use the following abridged notations:
$$\begin{array}{cccccccc}
w_{+++++}=w_{1},&w_{++++-}=w_{2},&w_{+++-+}=w_{3},&w_{+++--}=w_{4}\\
w_{++-++}=w_{5},&w_{++-+-}=w_{6},&w_{++--+}=w_{7},&w_{++---}=w_{8}\\
w_{+-+++}=w_{9},&w_{+-++-}=w_{10},&w_{+-+-+}=w_{11},&w_{+-+--}=w_{12}\\
w_{+--++}=w_{13},&w_{+--+-}=w_{14},&w_{+---+}=w_{15},&w_{+----}=w_{16},
\end{array}$$
and similarly,
$$\begin{array}{cccccccc}
w_{-++++}=\w_{1},&w_{-+++-}=\w_{2},&w_{-++-+}=\w_{3},&w_{-++--}=\w_{4}\\
w_{-+-++}=\w_{5},&w_{-+-+-}=\w_{6},&w_{-+--+}=\w_{7},&w_{-+---}=\w_{8}\\
w_{--+++}=\w_{9},&w_{--++-}=\w_{10},&w_{--+-+}=\w_{11},&w_{--+--}=\w_{12}\\
w_{---++}=\w_{13},&w_{---+-}=\w_{14},&w_{----+}=\w_{15},&w_{-----}=\w_{16}.
\end{array}$$

The method then proceeds as in the one-dimensional case.
The identities imposed by translational invariance on the correlators are
\beqa
c_1=c_2=c_3=c_4=c_5,\\ c_{12}=c_{14},c_{13}=c_{15},\
c_{25}=c_{34},\ c_{23}=c_{45},
\eeqa
where the indices $1,2,\ldots,5$, correspond respectively to the central spin, the East, North, West, South spins.
After reduction, the equations on the rates yield fourteen constraint equations. 
If one restricts the study to situations where the rates are invariant by spin symmetry, then the following additional constraints must be taken into account:
\beqa
\w_{1}=w_{16},\w_{2}=w_{15},\w_{3}=w_{14},\w_{4}=w_{13},\nonumber \\\w_{5}=w_{12},\w_{6}=w_{11},
\w_{7}=w_{10},\w_{8}=w_{9},\nonumber \\\w_{9}=w_{8},
\w_{10}=w_{7},\w_{11}=w_{6},\w_{12}=w_{5},\nonumber \\
\w_{13}=w_{4},\w_{14}=w_{3},\w_{15}=w_{2},\w_{16}=w_{1},
\eeqa
which in turn reduce the former system of fourteen constraint equations to a system of six equations, as follows:
\beqa\label{six}
\e^{8 K}\, w_{1}-\w_{1}=0,\nonumber \\
w_{6}-\w_{6}=0,\nonumber \\
\e^{4 K}\, w_{2}-\w_{2}+\e^{4 K}\, w_{5}-\w_{5}=0,\nonumber \\
\e^{4 K}\, w_{3}-\w_{3}-(w_{8}-\e^{4 K}\, \w_{8})=0,\nonumber \\
\e^{4 K}\,w_2-\w_2-(\e^{4 K}\,w_3-\w_3)
+\frac{2\e^{4 K}}{1+\e^{4 K}}(w_{7}-\w_{7})=0,\nonumber\\
\e^{4 K}\,w_2-\w_2+\e^{4 K}\,w_3-\w_3-\frac{2\e^{4 K}}{1+\e^{4 K}}(w_{4}-\w_{4})=0.\nonumber
\eeqa
The first two equations involve couples of rates related by detailed balance conditions, the following ones linear combinations of such couples. 
The rates now depend on ten independent parameters.

\subsection{Symmetric dynamics on the square lattice}

The dynamics is spatially symmetric, or undirected, if the left and right spins, or up and down spins have an equal influence on the central spin. 
For simplicity, we restrict the study to the most symmetric dynamics where the rates only depend on the number of up and down neighbouring spins of the central spin, i.e., rates which only depend on the value of the sum $\s_E+\s_N+\s_W+\s_S$:
\beqa\label{sym2d}
w_2=w_3=w_5=w_9=y,\nonumber\\
w_4=w_6=w_7=w_{10}=w_{11}=w_{13}=z,\nonumber\\
w_8=w_{12}=w_{14}=w_{15}=t.
\eeqa
Carried in the constraint equations above, we find
\beq\label{db2D}
t=\e^{4 K}\,y,\qquad w_{16}=\e^{8 K}w_1,
\eeq
and $z$ remains arbitrary. 
After fixing the overall scale of time the transition rates still depend on two independent parameters.
Fixing these parameters leads to the following simplified forms of the rates, which are simple generalisations of the 1D expressions,
\beqa
w({\s\s_E\s_N\s_W\s_S})
&=\frac{1}{2}\left(1-\s\tanh( K(\s_E+\s_N+\s_W+\s_S))\right),
\\
w({\s\s_E\s_N\s_W\s_S})&=\min\left(1,\e^{-\beta\Delta E}\right),
\\
w({\s\s_E\s_N\s_W\s_S})&=\e^{-\beta\,\Delta E/2}=\e^{-K\s(\s_E+\s_N+\s_W+\s_S)}.
\eeqa

\subsection{Directed dynamics on the square lattice}

Consider first the case where the central spin is only influenced by the North and East spins. 
This means that
\beqa
w_{1}=w_{2}=w_{3}=w_{4}=x,\nonumber\\
w_{5}=w_{6}=w_{7}=w_{8}= z,\nonumber\\
w_{9}=w_{10}=w_{11}=w_{12}=z',\nonumber\\
w_{13}=w_{14}=w_{15}=w_{16}=y
.
\eeqa
Carrying these relationships in the constraint equations~(\ref{six}) leads to 
\beq\label{ne}
y=\e^{8 K}\,x,\qquad z'=z,\qquad z=\e^{4 K}\,x.
\eeq
Out of the four unknown $x$, $y$, $z$, $z'$, only one remains undetermined, i.e.,
there only remains one free parameter in the expression of the rates.
The general expression of the rates satisfying these constraints reads
\beq\label{rate:ne}
w({\s \,\s_{E} \,\s_{N}\,\s_{W}\,\s_{S}})=\frac{\alpha}{2}
\left(1+\gamma^2\,\s_E\s_N-\gamma\s(\s_E+\s_N)\right),
\eeq
which is the {\it unique} solution of the question posed, up to the global timescale $\alpha$.
Fixing this timescale by the choice $\alpha=2\cosh^2 2 K$, allows to recast~(\ref{rate:ne}) into the compact form:
\beq\label{ku}
w({\s \,\s_{E} \,\s_{N}\,\s_{W}\,\s_{S}})
=\e^{-2 K\s(\s_{E}+\s_{N})}.
\eeq
Such rates violate detailed balance, but yet, as for the one-dimensional case, lead to a stationary state with Gibbs measure.
This form of the rates for the two-dimensional directed Ising model on the square lattice where each spin sees only its North and East neighbours was first proposed in~\cite{k} (up to a seemingly missing factor 2 in the exponent). 

We also considered the more general case where the central spin is influenced by three of its neighbours, say, West, North, East. Introducing the simplifying notations $u_1,\ldots, u_8$, we have
\beqa
w_{1}=w_{2}=u_{1},w_{3}=w_{4}=u_{2},w_{5}=w_{6}=u_{3},w_{7}=w_{8}=u_{4},\nonumber\\
w_{9}=w_{10}=u_{5},w_{11}=w_{12}=u_{6},w_{13}=w_{14}=u_{7},w_{15}=w_{16}=u_{8},
\eeqa
which carried into the six constraint equations~(\ref{six}) yield six new constraints
\beqa\label{wne}
u_{3} = \e^{4 K} u_{1},
u_{4} = \e^{4 K} u_{2},
u_{5} = (1+\e^{4 K}) u_{1} - u_{2},\nonumber\\
u_{6} = u_{3},
u_{7} = \e^{4 K} u_{5},
u_{8} = \e^{8 K} u_{1}.
\eeqa
The general expression of the rates satisfying these constraints depend on two independent parameters:
\beqa\label{rate:wne0}
w({\s \,\s_{E} \,\s_{N}\,\s_{W}\,\s_{S}})=\\\frac{\alpha}{2}
\left(1+c\,\s_W\s_N+(\gamma^2-c)\s_E\s_N
-\s\left(\frac{c}{\gamma}\s_W+\gamma\s_N+\bigg(\gamma-\frac{c}{\gamma}\bigg)\s_E\right)
\right).\nonumber
\eeqa
Let us now consider two particular cases of this general expression.
\begin{itemize}

\item
The previous case where the central spin is influenced only by the North and East spins is a particular case of the model just considered. In order to recover the results~(\ref{ne}) and (\ref{rate:ne}) it suffices to impose the additional conditions
\beq
u_1=u_2=x,\quad u_3=u_4=z,\quad u_5=u_6=z',\quad u_7=u_8=y
\eeq
in (\ref{wne}), yielding $c=0$ in~(\ref{rate:wne0}), that is~(\ref{rate:ne}).

\item
If we impose the left-right spatial symmetry on the rates~(\ref{rate:wne0}), i.e., if
 $u_2=u_5$, which itself leads to $u_4=u_7$,
then we obtain the {\it unique} solution
\beqa\label{rate:wne}
w({\s \s_{E} \s_{N}\s_{W}\s_{S}})=\frac{\alpha}{2}
\left(1+\frac{\gamma^2}{2}\s_N(\s_E+\s_W)-\frac{\gamma}{2}\s(\s_W+2\s_N+\s_E)
\right),\nonumber\\
\eeqa
up to the global scale $\alpha$.
This expression cannot be recast in an exponential form analogous to~(\ref{ku}).

\end{itemize}

\subsection{Directed dynamics on the triangular lattice}
We now consider the two-dimensional Ising model on the triangular lattice.
There are 128 ($2^7$) unknown rates, corresponding to the 128 motifs
where the central spin is surrounded by its six neighbours.
Using the same method as above, we obtain the set of constraints between the transition rates that need to be satisfied in order for the stationary state measure to be Gibbsian. 
The result is rather lengthy and will not be written here.

Let us specialize to the directed model
in which the subset of influential spins are any three consecutive spins at $\pi/3$ one from the other. 
The sites corresponding to the subset of influential spins and the sites corresponding to the complementary subset of neighbouring spins are thus related by spatial parity with respect to the central site.
For this directed case, we find a {\it unique} solution, up to a global timescale, to the problem of the determination of rates leading to a Gibbsian stationary state. 
The solution found has again the exponential form
\beq\label{tri}
w(\s\s_1\s_2\ldots\s_6)=\e^{-2K\s(\s_1+\s_2+\s_3)},
\eeq
where $\s_1$, $\s_2$ and $\s_3$ are the spins felt by the central spin, denoted by $\s$.

\subsection{Directed dynamics on the cubic lattice}

We finally consider the three-dimensional Ising model on the cubic lattice.
Again there are 128 ($2^7$) unknown rates, corresponding to the 128 motifs
where the central spin is surrounded by its six neighbours.
We find that the set of constraints between the transition rates that need
to be satisfied in order for the stationary state measure to be Gibbsian
are the detailed balance relationships.
In other words, for the three-dimensional Ising model on the cubic lattice, the only possible dynamics leading to the Boltzman-Gibbs distribution
at stationarity is the symmetric one.

\medskip
The method used in this section is unfortunately cumbersome when applied to lattices of increasing coordination number. 
The next section provides an alternate approach.

\section{Lattices with given coordination number}
\label{coordi}

In this section we introduce a complementary viewpoint on the general problem of the existence of rates leading to a Gibbsian stationary state, when the dynamics is directed. We proceed in two steps. First, we check the results of section~\ref{regular} for exponential rates: assuming the form of the rates, we show that they satisfy, or do not satisfy, the master equation at stationarity, depending on the lattice considered. Secondly, generalizing these observations, we give a method for the determination of which lattice, of given coordination number, is able or unable to sustain exponential rates, for directed dynamics.

\subsection{Check of the results of section~\ref{regular}}
The idea stems from the observation that, once the form of the rates satisfying the master equation~(\ref{mast}) have been determined by the method of section~\ref{regular}, it should be an easy matter to check the reverse, i.e., that the stationary master equation~(\ref{mast}) is satisfied.
For example, for the directed Ising chain with rates~(\ref{ku1D}), eq.~(\ref{mast}) leads to the following condition, for a given fixed configuration $\c$,
\beq
\sum_i\e^{-2K\s_i \s_{i+1}}=\sum_i\e^{-2K\s_{i-1} \s_{i}},
\eeq
which indeed holds, due to the translational symmetry of the system.

We now consider the general case of a lattice of given coordination number $z$.
We assume that the rates have the exponential form of eqs.~(\ref{ku1D}),~(\ref{ku}) and~(\ref{tri}),
\beq\label{directed}
w(\c_i|\c)=\e^{-2\beta\s_i h^+_i},
\eeq
where we denoted by $h^+_i$ the field felt by spin $i$ due to the subset $v^+(i)$ of its influential neighbours.
Likewise we denote by $h^-_i$ the field due to the complementary 
subset $v^-(i)$ of neighbouring spins.
We only consider the case where the subsets $v^+(i)$ and $v^-(i)$ are related by spatial parity with respect to the central spin. 
With the hypothesis~(\ref{directed}), the master equation~(\ref{mast}) gives
\beq
\sum_i\e^{-2\beta\s_i h^+_i}=\sum_i\e^{-2\beta\s_i h^-_i},
\eeq
or
\beq
\sum_i\prod_{j\in v^+(i)}\e^{-2K\s_i \s_j}=\sum_i\prod_{j\in v^-(i)}\e^{-2K\s_i \s_j},
\eeq
that is, with $\tau=-\tanh 2K$,
\beq\label{voisin}
\sum_i\prod_{j\in v^+(i)}(1+\tau\s_i \s_j)=\sum_i\prod_{j\in v^-(i)}(1+\tau\s_i \s_j).
\eeq
This equation must be satisfied order by order in $\tau$.
This leads to a set of geometrical constraints that a given lattice, of given coordination number, should satisfy, in order for the master equation to be satisfied.

Before using the method in all its generality we first check that for the 2D square lattice, the 2D triangular lattice and the 3D cubic lattice, we recover the results found in section~\ref{regular}: the constraints are satisfied for the two first lattices, not for the last one, as we now show.

We start with the 2D square lattice. We have $h^+_i=J(\s_E+\s_N)$ and $h^-_i=J(\s_W+\s_S)$.
The constraint equation~(\ref{voisin}) reads
\beq
\sum_i 1+\tau \s_i(\s_E+\s_N)+\tau^2 \s_E\s_N=\sum_i 1+\tau \s_i(\s_W+\s_S)+\tau^2 \s_W\s_S,
\eeq
which should be fulfilled for any given fixed configuration $\c$.
At order $\tau$ this imposes that the equation 
\beq
\sum_i \s_i(\s_E+\s_N)= \sum_i \s_i(\s_W+\s_S)
\eeq
be satisfied. This is obviously the case because of translation invariance.
At order $\tau^2$ the constraint reads
\beq
\sum_i \s_E\s_N= \sum_i\s_W\s_S,
\eeq
which again is easily seen to hold.
This completes the proof for the 2D square lattice.

We proceed with the 2D triangular lattice, with $z=6$. Let us denote by $e_1,e_2,e_3$ the unit vectors spanning the lattice. Then, at order $\tau$, the constraint equation reads
\beq
\sum_i \s_i\s_{i+e_1}+\s_i\s_{i+e_2}+\s_i\s_{i+e_3}=\sum_i \s_i\s_{i+\bar e_1}+\s_i\s_{i+\bar e_2}+\s_i\s_{i+\bar e_3},
\eeq
where $\bar e_1=-e_1$, and so on.
Again, trivially it is possible to pair the terms of both sides by translation.
At order $\tau^2$ the constraint is
\beqa
\sum_i \s_{i+e_1}\s_{i+e_2}+\s_{i+e_1}\s_{i+e_3}+\s_{i+e_2}\s_{i+e_3}
=\\
\sum_i \s_{i+\bar e_1}\s_{i+ \bar e_2}+\s_{i+\bar e_1}\s_{i+\bar e_3}+\s_{i+\bar e_2}\s_{i+\bar e_3}.
\eeqa
It is also easy to check that this equation is satisfied by translation invariance by pairing adequately the terms on both sides.
At order $\tau^3$, the constraint equation reads
\beq
\sum_i \s_i\s_{i+e_1}\s_{i+e_2}\s_{i+e_3}=\sum_i \s_i\s_{i+\bar e_1}\s_{i+\bar e_2}\s_{i+\bar e_3}.
\eeq
This equation is indeed satisfied for the triangular lattice because the following geometrical relationship between unit vectors holds:
\beq\label{geom}
e_2=e_1+e_3.
\eeq
Thus the quadrilateral formed by the spins on the left-hand side is related by translation to that appearing in the right-hand side.
This completes the proof for the 2D triangular lattice.

Let us now consider the 3D cubic lattice, with $z=6$. The constraint equations are the same as for the 2D triangular lattice, since these equations only depend on the coordination number and not on the structure of the lattice.
In the present case, the constraints are satisfied at order $\tau$ and $\tau^2$, but not at order $\tau^3$ since (\ref{geom}) no longer holds. We thus confirm the result of section~\ref{regular}, namely that the 3D cubic lattice cannot sustain exponential rates for directed dynamics.

\subsection{Which lattice of given coordination number can sustain exponential rates for directed dynamics?}

As exemplified by the two last cases considered above, where $z=6$, the constraints coming from equation~(\ref{voisin}) only depend on the coordination number $z$ of the lattice. Then, in all generality, for $z$ given, the question is whether there exists a lattice fulfilling these constraints or not. 

Consider for example the case of a lattice of coordination $z=8$.
The constraints at order $\tau$ and $\tau^2$ are trivially satisfied, as above. At order $\tau^3$ the constraint imposes, between the four unit vectors spanning the lattice, equalities of the type (\ref{geom}). At order $\tau^4$ the constraint imposes the relationship: $e_1+e_4=e_2+e_3$. This relationship can be implemented for the 3D centered cubic lattice, for example. However the constraints imposed at order $\tau^3$ cannot be fulfilled either by the 3D centered cubic lattice or by any other one.

Some more work can convince the reader that any lattice with coordination number $z\ge 8$ does not satisfy the constraints, hence cannot sustain a directed model with rates~(\ref{directed}). 

This method assumes rates of the form~(\ref{directed}).
It is however plausible that if these rates do not produce stationary states described by the Gibbs measure, then no other form is able to do so,
because the exponential form~(\ref{directed}) is the best fitted to be a solution of the master equation~(\ref{mast}).
We are therefore led to conclude that directed models with Gibbsian stationary states only exist in dimension one and two.

\section{Directed Ising-Glauber models}
\label{glob}

In this section we investigate
whether directed Ising models, augmented 
by Glauber dynamics, can exhibit a phase transition to a ferromagnetic state.
\subsection {The one- and two-dimensional directed Glauber-Ising models}
We begin by presenting the transition rates for directed Ising models with 
Glauber dynamics. 
Let $h^+_i$ denote, as above, the field felt by spin $i$ due to its influential neighbours $v^+(i)$.
For example, for the directed Ising chain
where each spin only sees the spin to its right, 
the field felt by spin $i$ due to spin $i+1$ is $h^+_i=J\sigma_{i+1}$. 
In the two-dimensional case where the spin $\sigma_{i}$ sees the spins 
$\sigma_{N}$ and $\sigma_{E}$, the field felt by spin $i$ is given by 
$h^+_{i} = J(\sigma_{N}+\sigma_{E})$. 
We define the Glauber rate for the process in which spin $\s_i$ 
is flipped in configuration
$\c=\{\s_1,\ldots,\s_{i},\ldots,\s_{N}\}$ resulting in configuration 
$\c_i=\{\s_1,\ldots,-\s_{i},\ldots,\s_{N}\}$, as
\begin{equation}\label{glauber}
w(\c_i|\c)= \frac{1}{2}(1 - \sigma_i \tanh \beta h^+_i).
\end{equation}

We now consider the equation of motion for the magnetisation, 
$m_i = \langle \sigma_i \rangle$, of the spin at site $i$,
first for the directed Ising chain with periodic boundary conditions. 
This equation reads (where a dot indicates a time derivative)
\begin{eqnarray} 
\dot{m}_i & = & -m_i + \langle \tanh (K\sigma_{i+1}) \rangle \nonumber \\
 & = & -m_i + \tanh K\ m_{i+1},
\label{Glauber1D}
\end{eqnarray}
($K=\beta J$).
The periodic boundary conditions, combined with translational invariance, 
imply that $m_i=m$, independent of $i$, giving 
\begin{equation}
m(t) = m(0)\exp[-(1-\tanh K)t].
\end{equation}
This shows that the one-dimensional system cannot support an ordered phase, 
since an initial magnetisation will decay exponentially to zero on time 
scale $t_{\rm eq} = (1-\tanh K)^{-1}$. 
Note that the stationary measure for this case is the Boltzmann distribution with temperature doubled, as can be seen from the alternate form of the rate~(\ref{alternate}).
Therefore the vanishing of the magnetisation in the stationary state of the 1D directed 
Ising model is not very surprising. 
However, a very similar analysis can 
be employed for the 2D directed Ising model defined above.

Let us label the sites with 
a pair of integers $(i,j)$, these being the coordinates in the two spatial 
directions, and suppose the site $i,j$ is influenced only the spins at 
sites $(i+1,j)$ and $(i,j+1)$. Then the equation of motion for $m_{i,j}$ 
reads 
\begin{eqnarray}
\dot{m}_{i,j} & = & -m_{i,j} 
+ \langle \tanh[K(\sigma_{i,j+1} + \sigma_{i+1,j})] \rangle \nonumber \\
& = & -m_{i,j} + \frac{1}{2}\tanh(2K)(m_{i,j+1} + m_{i+1,j}),
\label{Glauber2D}
\end{eqnarray}
where the last line follows from the fact that $\sigma_{i,j+1} 
+ \sigma_{i+1,j}$ can take only three values, $-1$, $0$ and $1$. 
With periodic boundary conditions, translational invariance implies that 
$m_{ij}=m$, independent of the site indices $i,j$, giving
\begin{equation}
m(t) = m(0) \exp[-(1-\tanh 2K)t].
\end{equation}
It follows that the magnetization again vanishes in the stationary state. 

One can also treat the case of open boundaries, where translational
invariance no longer applies. In one dimension, for example, the
right-most spin, $\sigma_N$, retains it initial value for all
$t$. Without loss of generality, we can fix $\sigma_N = 1$. In the
 stationary state, eq.\ (\ref{Glauber1D}) then gives $m_i = (\tanh
K)^{N-i}$, so the local magnetisation decays exponentially as a
function of distance from the fixed spin $\sigma_N$, and the total
magnetisation per spin vanishes in the thermodynamic limit.

\subsection{Cayley Trees}
We now discuss the case of directed Ising models on Cayley trees, for
the case of Glauber dynamics. We take the influence to be directed
from the tips towards the root, i.e.\ each spin only sees the spins
that are further from the root.

We consider first a tree with branching ratio 2, i.e.\ coordination 
number 3. We assume that the initial condition is one in which all spins 
at a given level of the tree have an equivalent initial condition, e.g.\ all 
spins up, or all spins either up or down with probabilities that are the 
same for all the spins at that level. Then the Glauber equation for the 
magnetisation $m_n$ of a site at the $n$th level of the tree, counting 
the tips as level 1, is 
\begin{equation}
\dot{m_n} = -m_n + \langle \tanh K(S_1^{(n-1)} + S_2^{(n-1)})\rangle,
\label{q=2}
\end{equation}
where, for a given spin at level $n$, $S_1^{(n-1)}$ and $S_2^{(n-1)}$
are the two spins at the $(n-1)$th level that influence the given spin.
Then, as for the $d=2$ case, one has immediately
\begin{equation} 
\dot{m_n} = -m_n + \tanh (2K)\,m_{n-1}.
\end{equation}
In the stationary state, this gives, $m_n=\tanh (2K)\,m_{n-1}$. Under
iteration, the site magnetisation is driven to zero, i.e.\ the
interior of the tree remains unmagnetised, at any non-zero
temperature, even when the spins at the tips are completely aligned
(and recall that, if initially aligned, they will stay aligned
forever, as nothing can change the tip spins in the directed
model). We conclude that there is no phase transition to a
ferromagnetic state in the directed Cayley tree with branching ratio
2.

It is instructive to contrast this with the properties of the undirected Cayley 
tree. In this case we will partially align the tip spins by applying a 
magnetic field $h$ to these spins, and ask whether, in the equilibrium state, 
the induced magnetisation propagates into the interior. It is trivial to 
trace out the tip (i.e.\ level 1) spins to obtain a new model in which the 
new tip spins (now level 2) experience a field $h'$ given by 
\begin{equation}
H'=q \tanh^{-1} [\tanh H \tanh K], 
\end{equation}
where $q$ is the branching ratio, $H=h/T$, and $K=J/T$, 
with $T$ the temperature, now a physical variable. 
This recurrence relation has a trivial fixed point at $H=0$, which is stable 
for $q \tanh K < 1$. In this regime, the interior of the tree is 
unmagnetised. For $q \tanh K > 1$, however, the recurrence relation has 
a stable non-trivial fixed point, and spins in the interior of the tree have 
a non-zero mean value. These two regimes are separated by a critical 
coupling $K_c$ given by $\tanh K_c = 1/q$. We conclude that the undirected 
model has a phase transition for all $q>1$, but for $q=2$ the directed 
model does not. 

To explore further the directed models we consider the case $q=3$. Since 
each spin at the $n$th level is influenced by three spins at the $(n-1)$th 
level, the analogue of eq.\ (\ref{q=2}) is 
\begin{equation}
\dot{m}_n = -m_n + 
\langle \tanh K[S_1^{(n-1)} + S_2^{(n-1)} + S_3^{(n-1)}]\rangle,
\label{q=3}
\end{equation}
Exploiting the fact that $(S_1+S_2+S_3)$ can only take four values 
$(3,1,-1,-3)$, and its odd parity under $S_i \to -S_i$ for all $i=1,2,3$, 
one can write
\begin{eqnarray}
&& \langle \tanh K[S_1^{(n-1)} + S_2^{(n-1)} + S_3^{(n-1)}]\rangle 
\nonumber \\
&& = 3A\,m_{n-1} - B \langle S_1^{(n-1)}\,S_2^{(n-1)}\,S_3^{(n-1)}\rangle,
\end{eqnarray}
where
\begin{eqnarray}
A &=& [\tanh K + \tanh(3K)]/4\ , \nonumber \\
B &=& [3\tanh K - \tanh(3K)]/3\ .
\end{eqnarray} 
It is clear that spins at the same level are uncorrelated, since they are 
influenced by disjoint sets of spins at lower levels. It follows that 
$\langle S_1^{(n-1)}\,S_2^{(n-1)}\,S_3^{(n-1)}\rangle = m_{n-1}^3$. 
In the stationary state, therefore, eq.\ (\ref{q=3}) reduces to 
\begin{equation}
m_n = 3A\, m_{n-1} - B\,m_{n-1}^3\ .
\end{equation}
This recurrence relation has two fixed points: a trivial one $m^*=0$,
which is stable for $A<1/3$, and a nontrivial one $m^* =
[(3A-1)/B]^{1/2}$, which is stable for $A>1/3$. There is a critical
coupling $K_c$ obtained from the condition $A=1/3$, i.e.\ $\tanh K +
\tanh(3K) = 4/3$, with solution $K_c=0.4753269$. We deduce that the
magnetisation deep in the interior of the tree vanishes for $K<K_c$,
and is nonzero for $K>K_c$, i.e.\ there is a phase transition at
$K=K_c$.

Higher values of the branching ratio $q$ can be treated in a similar
way. In each case one finds a phase transition at a critical coupling
value $K_c$, which decreases with increasing $q$. For $q=4$, for
example, one finds $K_c=0.3102182$.

It is interesting that the smallest integer $q$ for which there is a
phase transition on the undirected tree is $q=2$, whereas on the
directed tree it is $q=3$. This is consistent with our observation
that the 2D Ising model on a directed square lattice, in which each
spin is influenced by only two other spins, has no phase transition
and raises the question as to whether the 2D Glauber-Ising model on a
triangular lattice, with each spin influenced by (say) three other
spins, would have a phase transition.

\section{Discussion and Summary}
\label{concl}
In this paper we have addressed two open questions. 
The first, is under what circumstances the stationary states 
of directed Ising models have the conventional Boltzmann-Gibbs measure. 
In such cases, 
there will be a conventional ferromagnetic phase transition identical to 
that of the undirected Ising model, for space dimensions $d\ge 2$. 
We have shown that there is a unique set of rates
for which the Boltzmann-Gibbs measure describes the stationary state in dimension 
$d=1$ and $d=2$, both for the square and triangular lattices (where in the latter models each spin sees half of its
neighbours).
We have also shown that for the cubic lattice in dimension $d=3$, the only set of rates satisfying the constraints of stationary Gibbs measure are those satisfying detailed balance.
In one and two dimensions it is actually quite simple to 
demonstrate that the rates~(\ref{directed}) do lead to stationary states with the 
Boltzmann-Gibbs measure by simply substituting this measure into the 
stationary master equation. The advantage of the present approach is that 
it shows that these rates are unique. 
We have generalized this procedure and used it as a method for the determination of which lattice is able to sustain transition rates for directed dynamics with stationary Gibbs measure.
We confirmed that the 2D triangular lattice is able to do so, while
the 3D cubic lattice is not.
More generally we claim that lattices of coordination number $z\ge8$ cannot
be directed, if their stationary measure is Gibbsian,
and conclude that directed models with such measure only exist in dimension one and two.

The second, related question, 
concerning the existence of phase transitions in directed
Glauber-Ising models, was inspired by the numerical studies presented
in \cite{stau}, where it was demonstrated that a directed
Glauber-Ising model on a square lattice, with each spin seeing only
its North and East neighbours, does not exhibit a phase transition to
a ferromagnetic state. Here we have provided a simple analytic proof
that there is no transition. We have also shown that directed Cayley
trees exhibit a phase transition when the branching ratio $q$
satisfies $q \ge 3$.

A number of open questions remain. 
A completion of the proof of the absence of rates
producing a Gibbsian stationary state for directed lattices of coordination $z\ge8$ would be desirable.
It would be worth studying the 2D hexagonal lattice
and see whether we can apply the method of sections~\ref{chain} and~\ref{regular} to this case.
It would also be interesting to compare the dynamical properties of the directed and undirected models studied in the present work in one and two dimensions, when both share the same Gibbsian stationary measure.

Secondly, the results on the Cayley
tree suggest that the coordination number plays an important role, and
raises the question as to whether a 2D directed Glauber-Ising model in
which each spin sees more that two neighbours (e.g. a square lattice
in which each spin sees its North, East and West neighbours, or a
triangular lattice) could have a phase transition. Another obvious
question concerns the directed 3D model, with each spin seeing its
``North'', ``East'' and ``Up'' neighbours. Does this model, with
Glauber dynamics, have a ferromagnetic phase transition? The data of
ref.~\cite{stau} suggest not. Unfortunately, an analytic study for
this case does not seem straightforward.

\ack It is a pleasure to thank J.M. Luck for very
interesting discussions, S. Prolhac for his help in Mathematica programming and J. Lebowitz for pointing
ref.~\cite{k} to us. We gratefully acknowledge the hospitality of the 
Isaac Newton Institute for Mathematical Sciences, Cambridge, where this 
work was begun. 

\appendix{}

\section{}
In this appendix we give the form of the rates for the 2D directed Ising models on the square lattice
with Hamiltonian defined with two coupling constants, $J_1$, $J_2$.

For the model where each spin sees its North and East neighbours,
eq.~(\ref{rate:ne}) is replaced by 
\beq
w({\s \,\s_{E} \,\s_{N}\,\s_{W}\,\s_{S}})=\frac{\alpha}{2}
\left(1+\gamma_1\gamma_2\,\s_E\s_N-\s(\gamma_1\s_E+\gamma_2\s_N)\right),
\eeq
where
$\gamma_{1,2}=\tanh 2K_{1,2}$,
which, after fixing the global timescale by the choice $\alpha=2\cosh 2 K_1\cosh 2 K_2$,
yields the unique solution:
\beq
w({\s \,\s_{E} \,\s_{N}\,\s_{W}\,\s_{S}})
=\e^{-2\s(J_1\s_{E}+J_2\s_{N})}.
\eeq

For the model where each spin sees its West, North and East neighbours,
eq.~(\ref{rate:wne}) is replaced by
\beqa
w({\s \,\s_{E} \,\s_{N}\,\s_{W}\,\s_{S}})=\frac{\alpha}{2}
\left(1+\frac{\gamma_1\gamma_2}{2}\,\s_N(\s_E+\s_W)
-\frac{\s}{2}(\gamma_1\s_W+2\gamma_2\s_N+\gamma_1\s_E)
\right).\nonumber
\eeqa

%


\section*{References}

\begin{thebibliography}{99}


\bibitem{zrp} For a review, see: Evans M R and Hanney T 2005 J. Phys. A {\bf 38} R195
\nonum
Godr\`eche C 2007 Lect. Notes Phys. {\bf 716} 261

\bibitem{asep} For a review, see: Blythe R A and Evans M R 2007
J. Phys. A {\bf 40} R333

\bibitem{kls}
Katz S Lebowitz J L and Spohn H 1983 Phys. Rev. B {\bf 28} 1655
\nonum
Katz S Lebowitz J L and Spohn H 1984 J. Stat. Phys. {\bf 34} 497


\bibitem{gl} Luck J M and Godr\`eche C 2006 J. Stat. Mech. P08009 

\bibitem{stau} Lima F W S and Stauffer D 2006 Physica A {\bf 359} 423

\bibitem{glau} Glauber R G 1963 J. Math. Phys. {\bf 4} 297

\bibitem{k} K\"{u}nsch H R 1984 Z.Wahrscheinlichkeitstheorie
verw. Gebiete {\bf 66} 407 



\end{thebibliography}
\end{document}